# Momentum Injection via Dielectric Barrier Discharge Actuators in Low-Speed External Flow

Anthony Tang[1], Benjamin Price[1], Alexander Mamishev[2], Alberto Aliseda[1], Igor Novosselov[1,3,*]

[1]Department of Mechanical Engineering, University of Washington, Seattle, U.S.A. 98195

[2]Department of Electrical and Computer Engineering, University of Washington, Seattle, U.S.A. 98195

[3]Institute for Nano-Engineered Systems, University of Washington, Seattle, U.S.A. 98195

## ABSTRACT

Dielectric barrier discharge (DBD) plasma actuators can generate a wall jet without moving parts through interaction between ionized and neutral molecules in an electric field. The coupling between electro-hydrodynamic, turbulence, and viscous effects in the flow boundary layer remains unclear and deserves careful investigation. We present an experimental investigation of momentum injection by DBD actuators in a $U_\infty = 5$ m/s and $U_\infty = 11$ m/s co-flow and counter-flow configuration over a range of $V_{AC} = 12$ kV - 19.5 kV peak-to-peak at a frequency of 2 kHz. In the co-flow configuration, the DBD actuator adds momentum to the boundary layer, similar to an electrohydrodynamic (EHD) jet in quiescent conditions. In the counter-flow configuration, flow separation is observed at free stream velocity $U_\infty = 5$ m/s. The momentum displacement in the counter-flow configuration is ~ 6x greater than EHD jet momentum in a quiescent environment. Both co-flow and counter-flow momentum injections show diminishing effects with increased external flow speed. This work highlights that the resulting flow pattern is not a simple superposition of the EHD jet and the free stream but is determined by a balance between the inertial, viscous and Coulombic forces of the EHD and the external flow. The velocity profiles and momentum characteristics can be used to validate numerical models and inform the design of DBD actuators for active flow control.

## 1. INTRODUCTION

Non-thermal plasma devices can offer strategies for flow control [1-7]. Amongst active flow control devices, plasma actuators have the potential to control a fluid system while staying silent, instantaneous, and compact [8-10]. Corona discharge or dielectric barrier discharge (DBD), actuators generate ions when the electric field exceeds the dielectric strength of the working fluid. The interaction between free ions, accelerated by E-field, the working fluid, and surfaces can be utilized in aerodynamic drag reduction [11-13], lift augmentation [10, 14], separation control [15, 16], and electric propulsion [17-20]. Despite their lower electromechanical efficiencies than corona-driven flow, DBD actuators are more stable and can be more effective at providing a consistent electro-hydrodynamic (EHD) forcing [4, 9]. The current DBD applications are limited to flow control at low-speed conditions due to their relatively weak EHD forces [17, 21, 22]. Many studies have explored the multiphysics phenomena to optimize the electrical and mechanical effects of the DBD [16, 23-29]. The previous work modeling corona EHD can be relevant [7, 25, 27, 30]; early numerical efforts developed simplified DBD models and found that EHD flow can be modeled through the interactions of positive and negative ion charges considered [9, 29, 31]. More recent work has employed modified versions of these simplified models to numerically explore the performance of DBD, such as the evolution of the velocity field due to the DBD jet [28].

[*] ivn@uw.edu

Most reports describing EHD – flow interaction are currently limited to analysis of electroconvective instabilities at very low Reynolds numbers. Electro-convection (EC) phenomenon was first reported by G. I. Taylor in 1966, describing cellular convection in the liquid droplet [32]. Since then, EC has been observed in other systems with the interaction of electric force with fluids. In nonequilibrium EHD systems [32-53], poorly conductive leaky dielectric fluid acquires unipolar charge injection at the surface interface in response to the electric field. In charge-neutral electrokinetic (EK) systems, EC is triggered by the electro-osmotic slip of electrolyte in the electric double layer at membrane surfaces [54-65]. In 3D shear flow, the EHD addition to crossflow results in the formation of streamwise rolling patterns as observed numerically [51, 66-68] and experimentally [69, 70]. The 2D and 3D flow analysis related to DBD jet momentum injection in the shear flow has not been reported. Thus, a mechanistic understanding of the interaction between discharge and fluid flow in the presence of an external flow is needed to inform the development of DBD actuators for active flow control.

To maximize the effects of DBD actuators, recent experimental work varied actuator geometries such as electrode shapes, number of electrodes, and their placement on an aerodynamic surface. However, most of the fundamental EHD studies were performed in a quiescent environment; these actuators have not been well studied in external flow conditions, especially at low to moderate velocities relevant to flow separation control [71, 72]. Several airfoil and small-scale unmanned aerial vehicle (UAV) studies have explored the ability of DBD actuators to manipulate lift and drag forces; however, these studies did not provide insight into the fluid flow field and the underlying physics responsible for the lift and drag changes [73, 74]. The two traditional external flow conditions include co-flow, when the jet direction is the same as the external flow, and counter-flow when the momentum injection is opposite to the external flow. *Pereira et al.* reported forces measurements during the co- and counter-flow DBD jet injection. The authors found that the total EHD thrust (or the difference in EHD jet thrust and the shear stress due to the surface) was identical for both a co-flow and counter-flow [75]. However, the range of freestream velocities (0 – 60 m/s) in increments of 10 m/s did not address the regime where the EHD jet velocity is similar to the external flow (0 – 10 m/s) [75]. Understanding underlying fluid dynamics could be addressed by taking the velocity profiles in the vicinity of the momentum injection. *Bernard et al.* reported the velocity profiles of DBD actuators in co-flow and found that the effects of the DBD jet diminished at higher external velocities; the authors did not investigate counter-flow conditions [76]. The scientific literature does not report experimental work characterizing velocity profiles in the counter-flow momentum injection by DBD jet.

Over the past decade, several studies have been conducted on DBD mounted to various airfoils [14, 77, 78], multi-element airfoils [79], flaps [80, 81], and full-scaled or near full-scaled aircraft [17, 21]. In all studies, the DBD actuator demonstrated an ability to change aerodynamic performance by increasing airfoil lift, decreasing drag, or changing pitching moment. In addition, multiple DBD array systems have been tested with moderate success; many of these studies found the need to balance the simultaneous interactions between jets acting in opposite directions [71, 82]. Counter-flow conditions potentially offer the instantaneous ability to more efficiently manipulate the boundary layer by increasing drag, decreasing lift, and changing the pitching moment on an aerodynamic surface.

This present study explores the effect of momentum injection via the DBD actuator at $U_\infty = 5$ m/s and $U_\infty = 11$ m/s external flow in co-flow and counter-flow. To control the momentum injection, the AC voltage was varied in the $V_{AC} = 12$ kV – 19.5 kV range, and the frequency was set at 2kHz. This work provides insight into the interaction between the DBD jet and the external flow over a flat plate, informing the potential placement of the actuator on the airfoil and providing data for numerical studies.

## 2. EXPERIMENTAL SETUP AND DIAGNOSTICS

Traditional metrics to characterize plasma actuators' performance include current, power consumption, forces on the surfaces, and DBD jet velocity. A current measurement shows the capacitive current, discharge current, and noise superposition. The capacitive current is filtered out or ignored because it corresponds to the transiently stored energy in the dielectric or air and not the energy used to accelerate the fluid [5]. On the other hand, the discharge current indicates the amount of charged species that can participate in the energy transfer to fluid motion. The discharge current comprises of numerous peaks in the positive-going cycle due to streamer propagation with the addition of glow discharge during the negative-going cycle [83]. Recent works have characterized the relationship between DBD discharge, capacitance, power consumption, and DBD actuator performance [84, 85]. High-resolution measurements have shown that both semi-cycles contribute to EHD force, and their relevant contributions are the topic of active scientific discussions [5, 86, 87]. Velocity measurements are often obtained by pitot tubes or particle imaging velocimetry (PIV); these measurements characterize momentum transferred from charged species to neutral molecules. While PIV measurements can capture an entire fluid field, integrating a PIV system with an external flow source can be challenging. In addition, it maintains the risk of the tracer particle interacting with the electric field. DBD wall similarity analysis was recently proposed with moderate success [88]; however, additional experimental data is needed to perform a robust nondimensional analysis.

### 2.1. Wind Tunnel

All measurements were conducted in an open-circuit wind tunnel with a 100mm x 100mm cross-sectional test section. The wind tunnel consists of an inlet section followed by a 1000 mm long section and a test section with the provisions for inserting the DBD actuator plate. The DBD actuator plate surface rests coincident and colinear with the bottom wall of the wind tunnel, see Figure 1. The inlet section comprises four 120 mm x 120 mm fans with inlet cowls, a large honeycomb screen, and a contraction cone. The contraction cone initially employs a 9:1 contraction ratio to the resulting 100mm x 100mm wind tunnel section constructed of plexiglass. An aluminum extrusions frame supports the wind tunnel section. The velocity measurements were performed using a custom glass pitot tube with a 0.4mm ID and 0.5mm OD, as described below. The boundary layer height ($\delta_{99}$) at $U_\infty$ = 5 m/s external flow was measured at the location of the actuator to be ~11.0 mm.

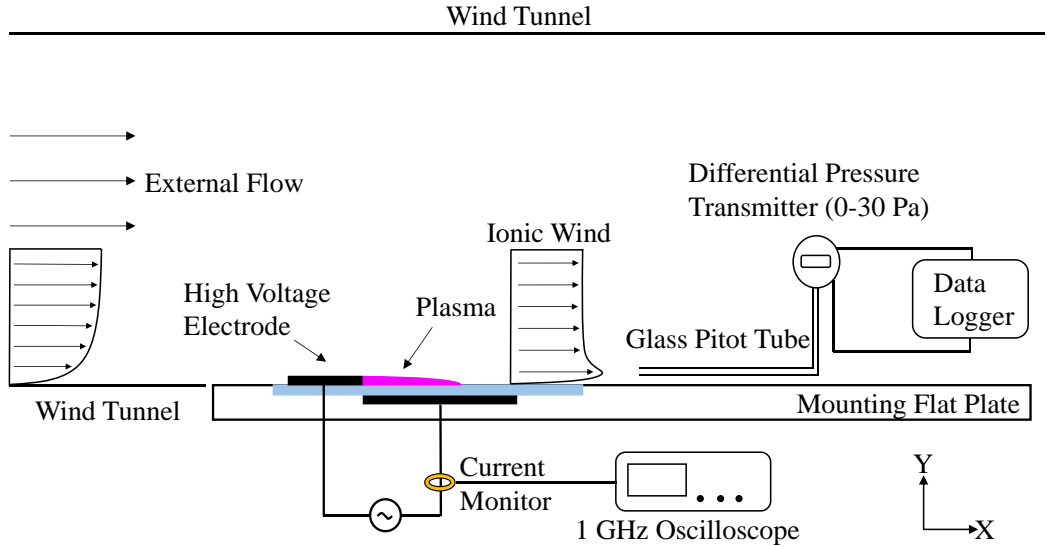

**Figure 1. Schematic of the experimental setup, DBD actuator is mounted on an acrylic glass plate flushed with the wind tunnel bottom. The blue region is the dielectric layer separating the electrodes.**

## 2.2. DBD actuator

The DBD actuator is comprised of two electrode DBD separated by a thin dielectric barrier, as shown in Figure 1, similar to the previous work [89]. When a high voltage is applied to the active electrode, the electric field is strongest at the edge of the active electrode, where the plasma is generated [86, 90, 91]. The downstream length of the electrodes is usually a few millimeters, and different studies have explored the effects of gaps between the electrodes [5, 92, 93]. The electrodes' thickness and the dielectric media can impact the actuator's performance [91, 94-97]. A straight-edged DBD actuator with a spanwise uniform electric field produces a two-dimensional forcing on the fluid, resulting in a planar jet. Other actuator designs have been considered, including serrated electrodes that produce a three-dimensional force in a flow field [98-100]. The spanwise length or width of the electrodes serves as a nominal reference length in the analysis [5]. The DBD actuator was installed on the 6″ by 8″ acrylic plate for this study. The dielectric material used in this study is Kapton (~3.5 dielectric constant at 1 kHz). Each actuator has 1 layer of 3 mil Kapton-FN (FEP layered Kapton) and 4 layers of 1 mil Kapton-HN with a total thickness of ~ 15 mil (including the adhesive and FEP layers). The ground electrode (copper, 50 µm thick, 25 mm long, 110 mm wide) is flush-mounted on the acrylic plate. The upper electrode (copper, 50 µm thick, 15 mm long, 110 mm wide) is glued onto the top of the Kapton dielectric layer. Both electrodes have straight edges producing a uniform spanwise discharge. The active and ground electrodes' edges are aligned with each other, i.e., there is no overlap between the electrodes in the x-direction. The air-exposed HV electrode is connected to a Trek 615-10 (PM04014) power supply that provides up to 20 kV (peak-to-peak) AC high voltage.

## 2.3. Electrical Measurements

The electric current in the DBD circuit is a superposition of a capacitive current and a discharge current. The discharge current is associated with plasma microdischarges, and they appear as a series of fast current pulses [101], as shown in Figure 2(a). DBD current is measured using a 200 MHz bandwidth non-intrusive Pearson 2877 current monitor with a rise time of 2 ns. The current monitor is connected to a Tektronix DPO 7054 oscilloscope that uses a bandwidth of 500 MHz to satisfy the Nyquist theorem by achieving a sampling rate of 1 GS/s. These conditions are essential for accurately capturing individual discharges that have been shown to occur on average over a 30 ns duration [102]. The high bandwidth and the sampling rate minimize the noise during the current measurements and can be used to compute the time-averaged electrical power [52]. The voltage from the power supply is also simultaneously displayed on the oscilloscope. To determine the currents associated with the plasma microdischarges, some have explored the capacitive current through analytical methods [103, 104]; others have removed the capacitive current through signal processing methods, including low-pass filters or Fast-Fourier Transform (FFT) [87, 101, 102]. For example, our previous work determined capacitive current by considering the first 15 Fourier modes. The discharge events appear as pulses emerging from a noisy baseline.

Here, to determine the power consumed by the actuator, a Lissajous curve is created by introducing a capacitor between the grounded electrode and the ground [105, 106]. The integration of the charge-voltage relationship multiplied by the frequency yields the total power usage of the DBD system. The time-averaged electrical power consumed by the actuator can be computed as

$$W_{elec} = f_{AC} \int_{t^*=0}^{t^*=1} V dQ, \qquad (1)$$

where $f_{AC}$ is the frequency of the applied voltage in Hz, and $V$ and $Q$ are respectively the voltage and charge at each point in the period. The normalized time ($t^*$) represents a single cycle. We compute the averaged resulting power from at least four separate periods to reduce the noise impact. In the wind tunnel study by *Pereira et al.,* the DBD actuator in co-flow and counter-flow was found not to have significantly different electrical characteristics [75]. Figure 2(a) below shows a typical DBD current measurement with a voltage curve. Figure 2(b) shows the representative filtered Lissajous curve of four consecutive discharge cycles. These data were used to determine the power of the DBD actuator as a function of the operating condition.

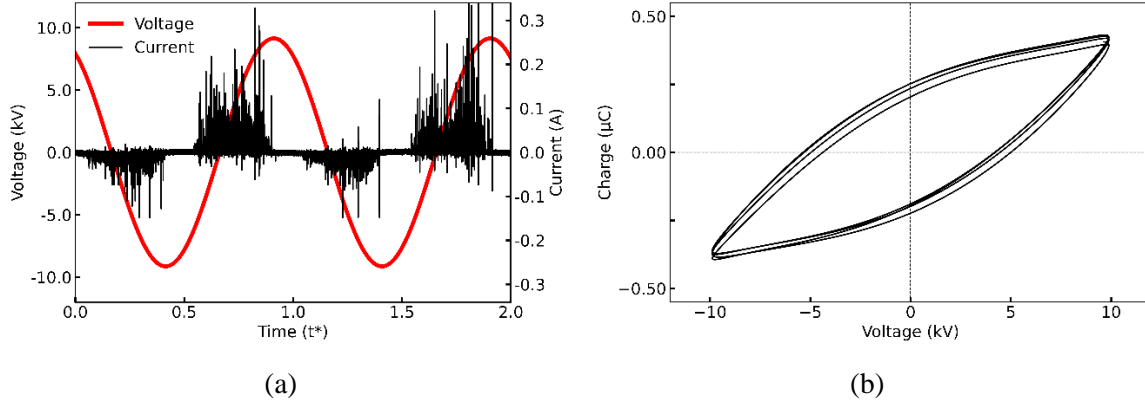

(a) (b)

**Figure 2. (a) Typical DBD current with voltage signals at 18 kV (p-p) and 2 kHz applied frequency (b) 4 consecutive Q-V discharge cycles measured from a 100 nF capacitor.**

### 2.4. Wall jet and momentum displacement characterization

The flow field induced by EHD characterizes the fluid mechanical properties of the plasma actuators. We employed a custom-made glass pitot tube with a 0.4 mm inner diameter and 0.5 mm outside diameter to measure the time-averaged velocity profile. Compared to traditional stainless steel pitot tubes, the glass tube reduces electrical interaction with the discharge. This method has been previously used to characterize plasma actuators' performance [5, 89, 101, 107]. The pitot tube is mounted on an optical table and controlled on the x and y-axis by linear stages connected to an Ashcroft CXLdp differential pressure transmitter (0 – 25 pascals with 0.25% accuracy). The pressure transducer outputs a 4 – 20 mA current linear to its pressure range and is in series with a 1.5 kΩ resistor. The pressure within the pitot tube converges nearly instantly. The voltage across the resistor is recorded for a minimum of 30 seconds across a Hydra Data Logger II. With the time-averaged pressure ($P$), a time-averaged wind velocity ($v$) is calculated using Bernoulli's equation with a calibration correction factor ($C$) that is characteristic to custom pitot tube expressed as

$$\Delta P = C\rho v^2, \qquad (2)$$

where $\rho$ is the fluid density. In our experiments, the typical velocity measurements had a standard deviation < 0.02 m s$^{-1}$ over a 30 s sampling period. Only x-velocity measurements are taken at varying x and y positions downstream and upstream on the active electrode edge in this experiment. At each x-position, the y-velocity profile is obtained from the surface to 10 mm above the plate at increments of 0.25 mm or 0.5 mm (at a higher location). The streamwise measurements were taken by holding a constant y-position and spaning in the x-direction at 0.5 mm intervals to complete a grid data set. Due to the pitot tube dimension, we could not capture velocity at y < 0.25 mm. We assume the velocity is linear between the no-slip condition at y = 0 mm and the data at y = 0.25 mm for plotting purposes.

Considering 2D control volume with spanwise direction normalized by a unit dimension, from a vertical velocity profile of the wind tunnel with the DBD actuator, the total mass flow rate per meter spanwise, $Q$, of the system can be computed by

$$Q_{system} = \rho \int_{y=0}^{y=\infty} U(y) dy, \qquad (3)$$

where $U(y)$ is the measured velocity at varying heights at a constant x location. Similarly, the system's total momentum per meter spanwise can be found by multiplying the mass flow rate at each vertical position with its respective velocity such that

$$M_{system} = \rho \int_{y=0}^{y=\infty} U^2(y) dy. \tag{4}$$

To identify the momentum produced by the DBD actuator, the same measurements were taken without the DBD actuator OFF, and the difference in momentum should theoretically be the momentum produced by the actuator if viscous losses are neglected. The resulting momentum is expressed as

$$M_{DBD} = M_{system} - M_{wind\ tunnel}, \tag{5}$$

and this relationship similarly holds for mass flow rate and mechanical power. The mechanical power of the system ($W_{mech}$) can be computed by

$$W_{mech} = \frac{1}{2} \rho L \int_{y=0}^{y=\infty} U^3(y) dy. \tag{6}$$

The derived values of mass flow rate, momentum, and power of the DBD in external flow are compared to the results of similar measurements in quiescent flow.

## 3. RESULTS AND DISCUSSION

### 3.1. Power Consumption

Figure 3**Error! Reference source not found.** illustrates the power usage of the DBD for a range of operating voltages, external flow speed, and DBD actuator orientation. Integration of Lissajous Q-V curves yielded an average power consumption of the actuator. The power usage is normalized to the spanwise length of the actuator (0.11 m). In general, the power consumption increase quadratically with applied voltage, which is consistent with previous reports for AC DBD [5, 86, 89, 108] and the EHD flows driven by corona discharge [26, 30, 109]. Power usage between the AC cycles for any configuration had a maximum variance of ~ 8%, see Figure 3. These data were taken for the DBD actuator in quiescent, counter-flow, and co-flow conditions at the two external flow speeds. *Pereira et al.* also found power usage to vary less than 10% between co-flow and counter-flow forcing [75]. The magnitude of normalized power usage measured for this study is slightly higher compared to *Pereira et al.;* however other studies have found similar levels of power consumption and noted a range of power expenditures based on several different geometric parameters such as dielectric thickness [1, 5, 89].

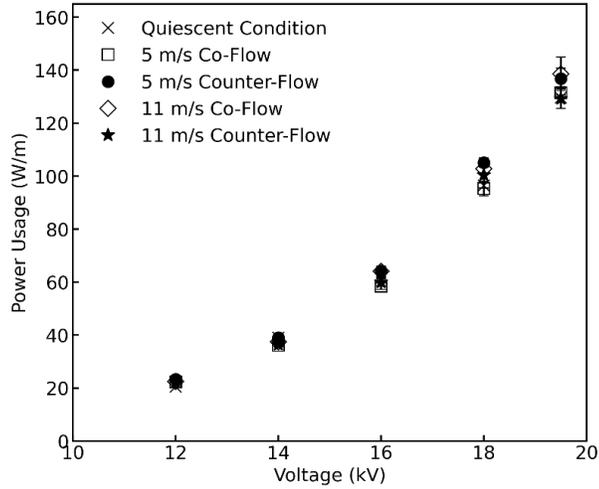

**Figure 3. DBD power usage at $U_\infty = 5$ and $U_\infty = 11$ m/s external flow in co-flow and counter-flow**

### 3.2. Operation in Quiescent Condition

Viscous effects in the boundary layer play an important role in momentum displacement from a DBD jet. Jet based on the momentum injection or impinging jet can be parameterized based on their similarity. The wall jet is typically divided into three regions: a self-similar wall layer where viscous forces are dominant, a self-similar outer layer that behaves analogously to a free jet, and an overlap layer with source dependence where the velocity is closest to the maximum. A triple-layered incomplete similarity is achieved by matching the self-similar outer and wall regions with the overlap layer [110-113], while at the wall, the jet decays due to viscous dissipation, shear effects in the outer layer, lead to flow entrainment. Figure 4 shows the velocity profile of the DBD jet at $x = 10$ mm and $x = 25$ mm downstream of the active electrode edge [89]. Though the velocity decay and the spreading of the jet are apparent as in the other wall jets, the presence of Coulombic forces associated with the DBD jet and the fact that the fluid is being accelerated over the volume of fluid rather than the point source make the parameterization of the jet extremely challenging. The addition of co- and counter-flow further complicates the analysis.

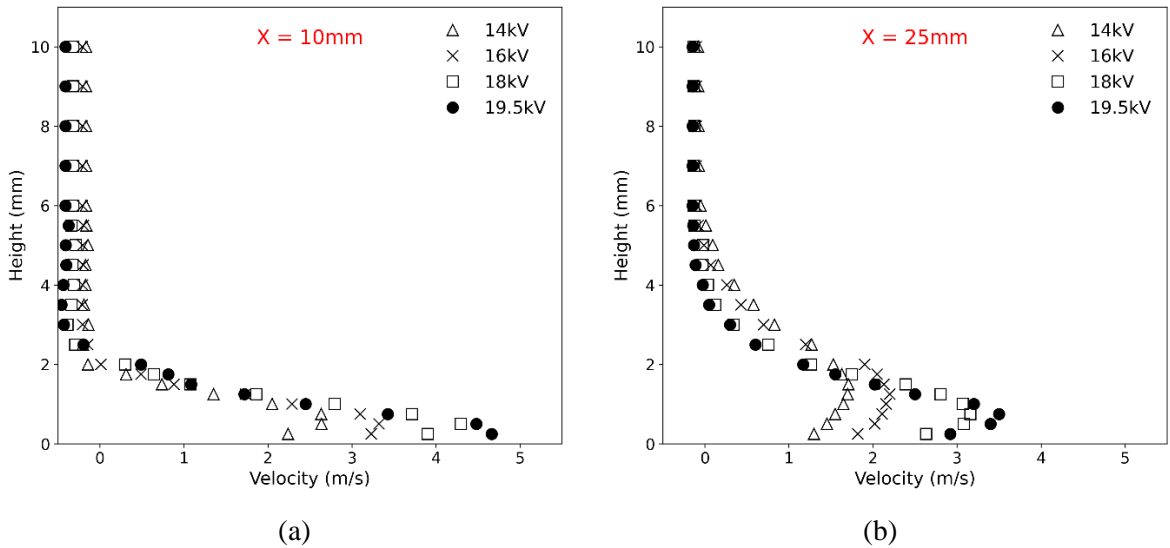

(a)　　　　　　　　　　　　　　(b)

**Figure 4. DBD with no external flow at $x = 10$ mm (a) and $x = 25$ mm (b) downstream. The maximum velocity at 19.5kV is approximate 4.7 m/s at a $y = 0.5$ mm at $x = 10$ mm downstream.**

### 3.3. Co-flow EHD Momentum Injection

This section discusses the DBD actuator performance in co-flow momentum injection over the range of $V_{AC}$ = 14 – 19.5 kV, frequency of 2 kHz for $U_\infty$ = 5 m/s and $U_\infty$ = 11 m/s. The virtual origin and coordinate system are defined in Figure 5. The velocity profiles of the EHD jet are measured at x = 10 mm and x = 25 mm downstream of the active electrode edge. Due to EHD momentum injection, in both external flow cases, the velocity in the boundary layer increases; however, the difference in momentum displacement shows the effect of boundary layer thickness.

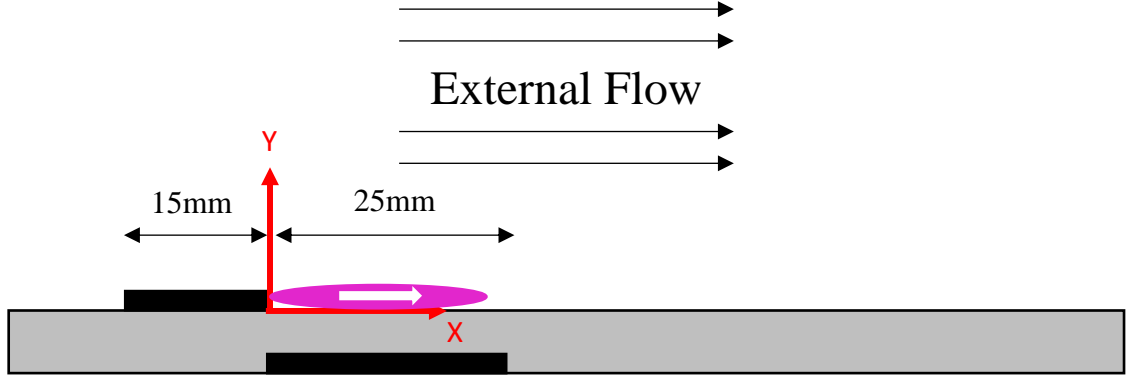

**Figure 5. DBD actuator in co-flow configuration, the first measurement is taken at x=10 mm to avoid plasma region disruption with pitot probe. The plasma region is colored in purple.**

Figure 6**Error! Reference source not found.** shows the velocity data at varied DBD voltage at two downstream locations. The dotted line is the wind tunnel velocity profile without plasma injection. Note that in quiescent conditions (Figure 4), the DBD jet has a maximum velocity, $U_{max}$ ~ 4.5 m/s at y = 0.25 mm at x = 10 mm downstream at $V_{AC}$ = 19.5 kV and $f_{AC}$ = 2 kHz, and the velocity of the jet after its maximum velocity diminishes quickly after its peak. Here, at $U_\infty$ = 5 m/s external flow, the EHD jet velocities are similar to that of the free stream, and the effects of the actuator are prominent. The increase in boundary layer velocity of ~ 2 m/s is nearly identical to that of F*Bernard et al.* [76] for a similar case. The $U_{max}$ ~ 5.4 m/s is greater than in quiescent condition (~ 4.8 m/s) but located at y = 0.75 mm and x = 10 mm at the same position. The DBD-induced momentum in co-flow does not dissipate as fast as in a quiescent environment, as seen at x = 25 mm. This can be explained by the interaction of the outer layer of the jet with the free stream, in which, instead of dissipating its momentum to the quiescent environment, the jet entrains momentum from the free stream. This entrainment is confirmed by a slight decrease in the external flow with the energized DBD actuator. It is expected that the fluid inside the wind tunnel control volume conserves mass, so adding momentum in the boundary layer leads to the decrease of free stream velocity. The scan of the wind tunnel profile confirmed that the fluid momentum is conserved.

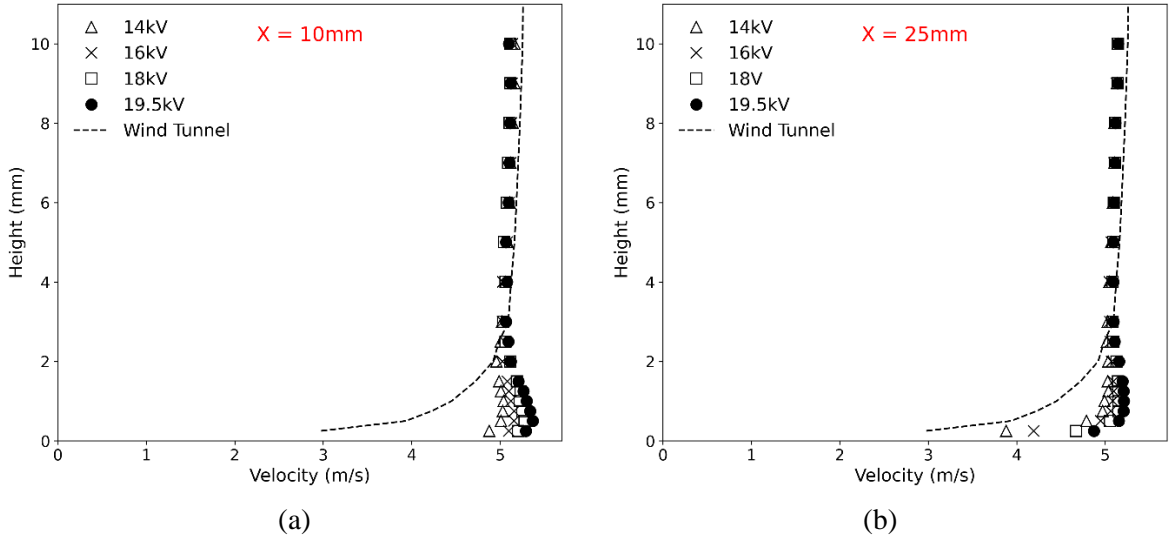

**Figure 6.** DBD actuator in $U_\infty$ = **5 m/s** co-flow at (a) **x = 10 mm** and (b) **x = 25 mm** downstream. The dashed line shows the freestream profile without plasma injection. The DBD voltage is varied in the 14kV-19.5kV range; the AC frequency is set constant at 2kHz.

Figure 7 shows the effect of the EHD jet in the boundary layer at freestream velocity at $U_\infty$ = 11 m/s. In this case, the EHD jet velocity is > 2x lower than the free stream flow for the highest DBD settings. The effects of the momentum injection are diminished, as the enhanced mixing in the thinner (than the 5 m/s case) boundary layer is more effective at dissipating EHD jet momentum. At maximum DBD power, the velocity increase is < 1 m/s at x = 10 mm. This locally thins the boundary layer; however, at x = 25 mm, the effect of the EHD momentum injection is negligible. These results agree with *Bernard et al.* [76] at $U_\infty$ = 10 m/s. At higher external flows, the EHD momentum addition results in lower momentum transfer rates, as the enhanced mixing in the thin boundary layer rapidly restores the boundary layer profile to the free stream shape.

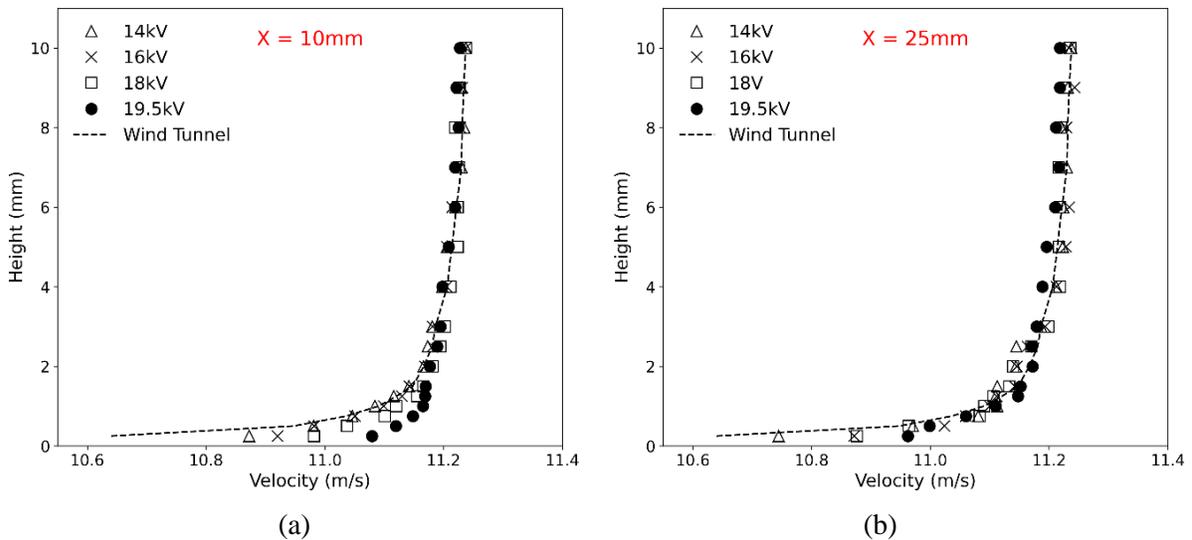

**Figure 7.** DBD actuator in $U\infty$ = **11 m/s** co-flow at **x = 10 mm** (a) and **x = 25 mm** (b) downstream. The DBD voltage is varied in the 14kV-19.5kV range, the AC frequency is set constant at 2kHz.

The EHD momentum addition cannot be treated as a linear superposition of the EHD jet in a quiescent environment and the momentum associated with the boundary layer of the free stream. For external flows compatible with EHD wall jet velocities, the momentum injection into the co-flow leads to the effective thinning of the boundary layer; this effect diminishes at higher freestream velocities. The wall jet dissipation is influenced by (i) interaction with freestream in the outer layer and (ii) viscous dissipation in the inner layer.

### 3.4. Counter–Flow EHD jet

This section characterizes the behavior of counter-flow EHD jet at DBD $V_{AC}$ = 14 kV – 19.5 kV at $f_{AC}$ = 2 kHz, and wind speeds of $U_\infty$ = 5 m/s and $U_\infty$ = 11 m/s. The virtual origin and coordinate system of the DBD in counter-flow are defined above in Figure 8. The datum for analysis is set at the plasma generation edge of the active electrode; however, the EHD momentum injection is now in the negative x-direction.

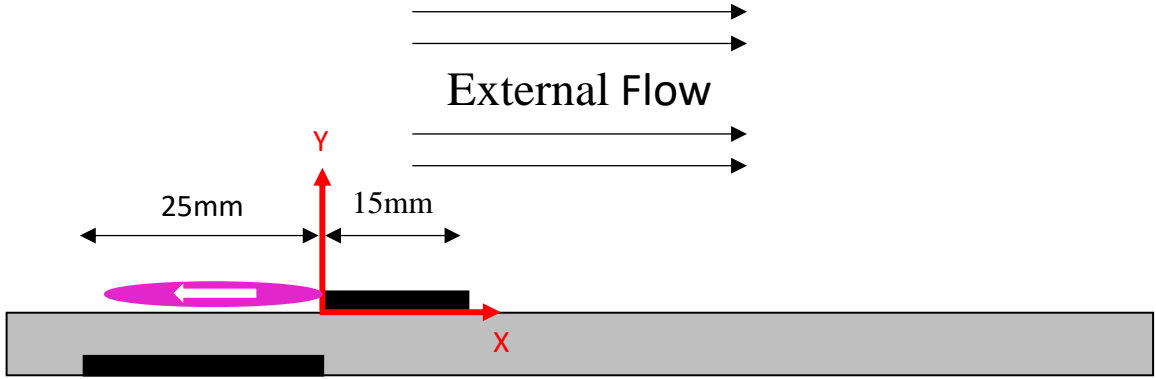

**Figure 8. DBD actuator in counter-flow configuration. The plasma region is colored in purple.**

Figure 9 and Figure 10 show the velocity profiles for the EHD momentum injection into counter-flow. The dotted line is the measured wind tunnel velocity profile with DBD actuator OFF. In the 5 m/s external flow case, the velocity of the EHD jet has a similar magnitude to the external flow resulting in an adverse pressure gradient and the formation of a recirculation zone. The exact boundaries of the separation region are difficult to determine experimentally in the plasma region (x < 0 mm, y < 2mm) as the insertion of the pitot probe into the plasma interferes with the experiment, see Figure 8. However, to preserve continuity, the EHD jet must entrain fluid from above and behind the jet; thus, the downstream y-scan and the x-scan at fixed y positions can be used to determine the boundaries of the separation bubble.

First, we examine the y-scan at the fixed x-position. Figure 9 and Figure 10 show the profiles at x = 10 mm (above the active electrode) and x = 25 mm for $U_\infty$ = 5 m/s and $U_\infty$ = 11 m/s, respectively. As in the co-flow experiments, the EHD jet strength is varied by varying $V_{AC}$ = 14 kV - 19.5 kV range; the AC frequency is held constant at 2kHz. For all voltages, the DBD in counter-flow displaces more momentum than its co-flow counterpart, e.g., the counter-flow EHD jet changes $U_{max}$ > 5 m/s at $V_{AC}$ = 19.5 kV compared to the ~ 2 m/s in the co-flow case. Note that the maximum negative velocity is likely located in or just after the EHD momentum injection region (x= - 10 mm – 0 mm). However, measurements could not be taken within the plasma region due to the plasma interactions with the pitot tube. Figure 9 shows that in the $V_{AC}$ = 19.5 kV case, the separation bubble extends past x = 10mm downstream of the active electrode edge, while other conditions exhibit flow reattachment. The flow is attached at x = 25 mm; however, the pressure gradient in the flow boundary layer has not yet recovered.

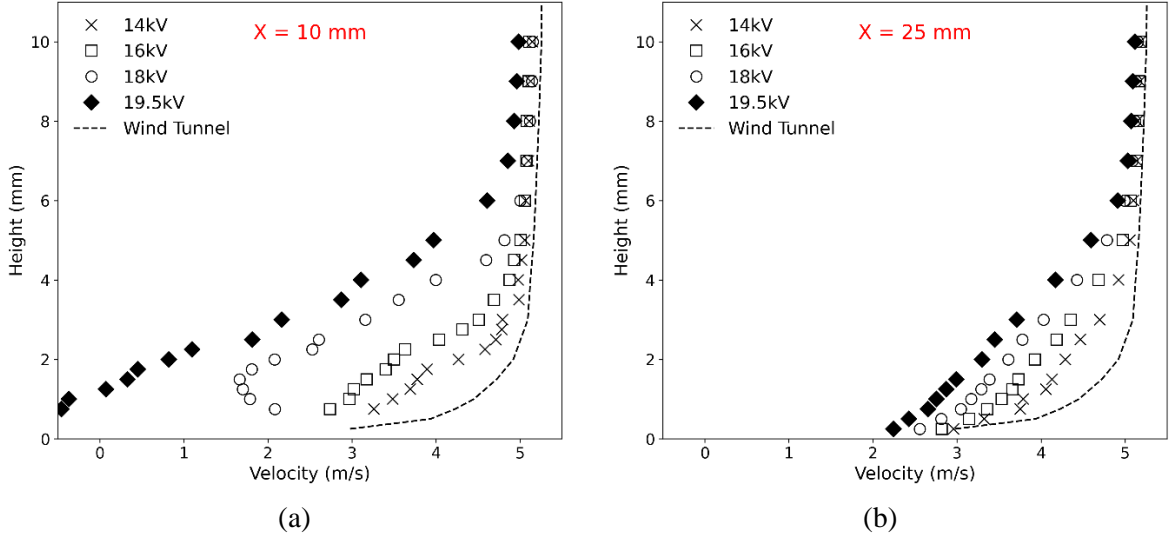

**Figure 9. DBD actuator in $U_\infty$ = 5m/s counter-flow at x = 10 mm (a) and x = 25 mm (b) downstream. The DBD voltage is varied in the 14kV-19.5kV range, the AC frequency is set constant at 2kHz.**

For stronger external flow, the effects of the DBD jet are diminished compared. Within the boundary layer, the largest decrease in velocity in counter-flow with $U_\infty$ = 11 m/s is approximately 2.0 m/s at y = 0.5 mm and x = 10 mm. No separation was observed in the $U_\infty$ = 11 m/s conditions, though the plasma region was not probed. While the effects of the DBD in counter-flow at $U_\infty$ = 11 m/s decrease compared to the slower free stream experiments, the effects of the EHD wall jet are more significant than in co-flow with the same $U_\infty$.

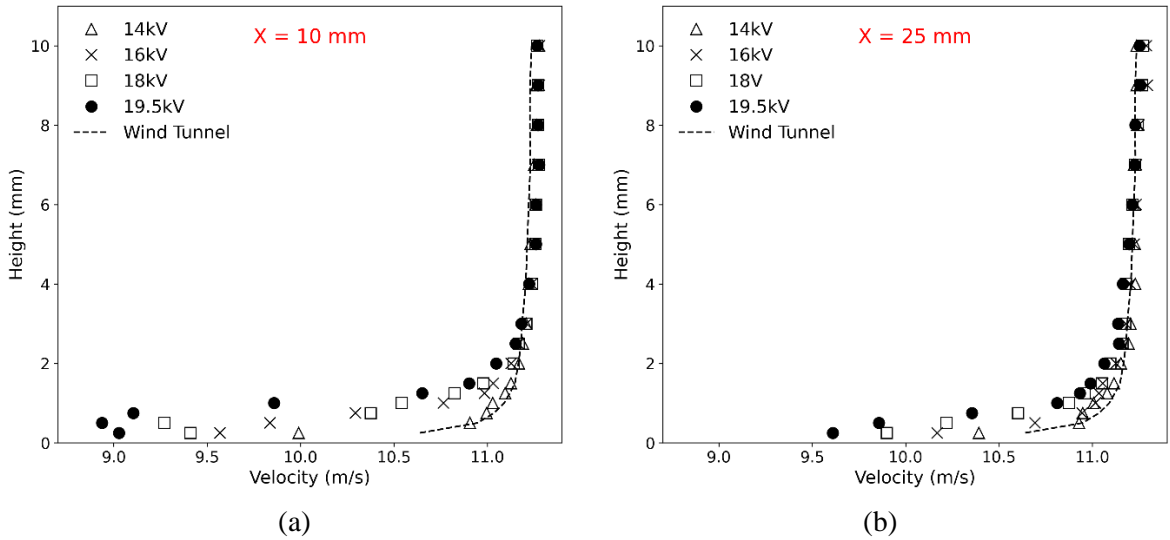

**Figure 10. DBD actuator in $U_\infty$ = 11 m/s counter-flow at x = 10 mm (a) and x = 25 mm (b) downstream. The DBD voltage is varied in the 14kV-19.5kV range, the AC frequency is set constant at 2kHz.**

The x-scans were performed holding the constant y position to determine the separation region boundaries. Figure 11 shows the velocity profiles for y = 0.5 mm and y = 1.0 mm, while the x position was varied from x = 0 mm (edge of the active electrode) to x = 15 mm. The DBD voltage was in the range of $V_{AC}$ = 12 kV - 18 kV at f = 2 kHz. The data for the condition of $V_{AC}$ = 19.5 kV is not shown due to the limited range of the pressure gauge.

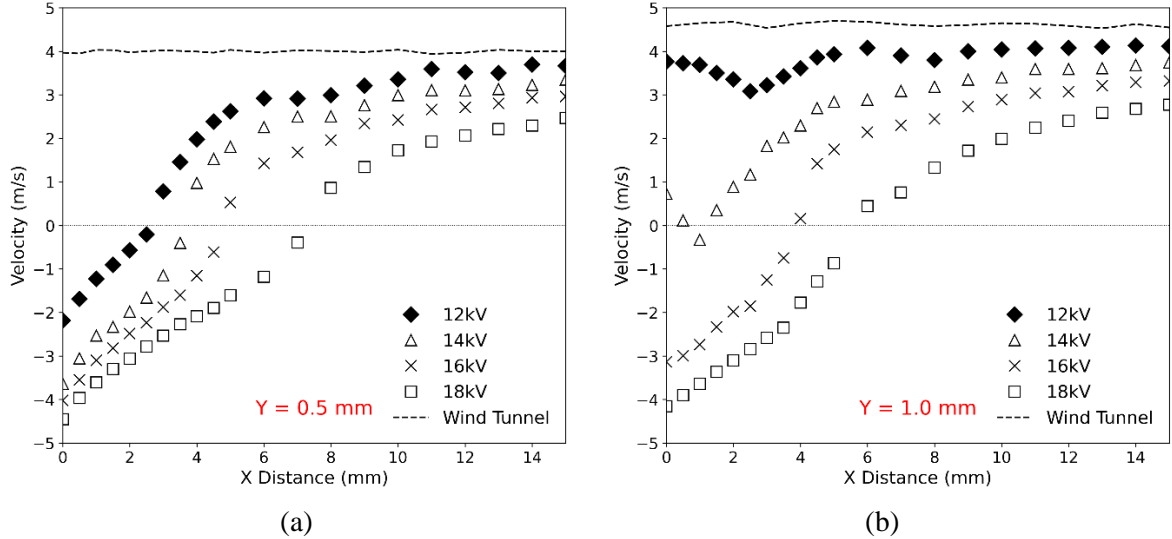

**Figure 11. DBD actuator in $U_\infty$ = 5 m/s counter-flow at y = 0.5 mm (a) and y = 1.0 mm (b). The DBD voltage is varied in the 14 kV-19.5 kV range, the AC frequency is set constant at 2 kHz.**

At y = 0.50 mm or immediately above the surface of the dielectric layer, separation is observed at all voltages. The x-location where the velocity vector changes from negative to positive determines the border of the separation bubble. At y = 1.0 mm, there is no separation at VAC = 12 kV; however, it exists at the other conditions. At VAC = 18 kV, the separation length is approximately 7.5 mm downstream.

To better visualize the flow pattern, additional x-scans were performed, and the 2D velocity fields were reconstructed. Figure 12 shows the velocity contour plots for counter-flow EHD jet obtained by merging x and y scans at $U_\infty$ = 5 m/s. Each grid point in the figure is associated with a velocity measurement; the spatial resolution was 0.5 mm in both x and y directions, totaling 200 measurements for each condition. All DBD conditions show a separation region with a negative x-velocity downstream of plasma injection. With the increase in DBD voltage, the separation extended from 3.0 mm (12 kV) to >10.0 mm (18 kV) in the x-direction from the edge of the active electrode and 0.6 mm (12kV) to > 1.75 mm (18 kV) in the y-direction. The growth in the length and height is nonlinear with increasing voltage. The size and shape of the recirculation bubble are determined by the competition of the EHD jet strength and direction of the jet vs. the local velocity in the boundary layer. As the EHD jet adds momentum at high DBD voltages, it can overcome the fluid velocity in the boundary layer at greater heights. Without sampling within the plasma region, it is challenging to characterize the entire length of the separation regions. It can be expected that the separation region extends into the forcing plasma region. Multiphysics CFD simulations can potentially address this issue; however, robust models need to be developed and validated. Though the CFD is beyond the scope of this paper, the data presented here can be used for such validations.

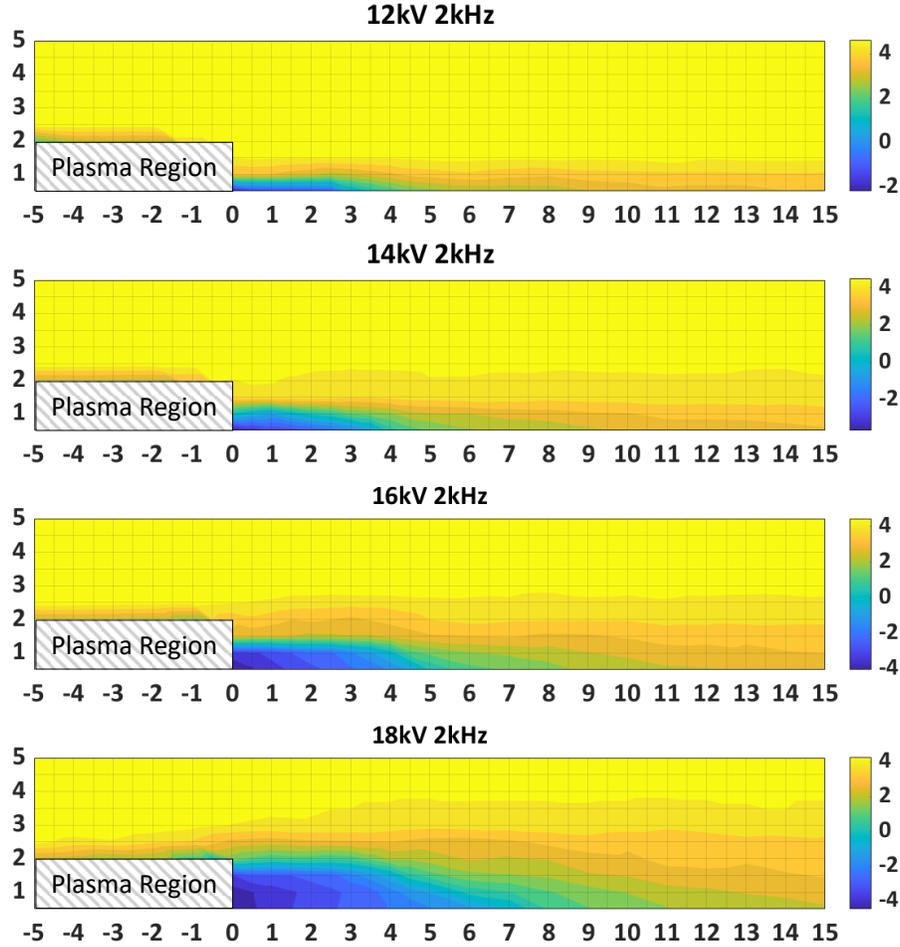

**Figure 12**. X-velocity contour plot for EHD jet in counter-flow for $U_\infty$ = 5 m/s for varying voltages. Gridlines correspond to recorded points spaced 0.5 mm apart in the x- and y-direction.

### 3.5. Momentum Displacement and Energy Transfer Characteristics

This section discusses the momentum displacement due to the EHD momentum injection at U$_\infty$ = 5 m/s and U$\infty$ = 11 m/s external flow. The momentum displacement is calculated by integrating the velocity profiles in the y-direction (y = 0 – 10 mm) at x = 10 mm downstream of the DBD wall jet. Note the x = 10 mm location is in the direction of the external flow with the datum at the plasma generation edge of the DBD actuator. Thus, in the co-flow case, the x = 10 mm location is downstream of the momentum injection; however, in the counter-flow case, the x = 10 mm location is upstream of the actuator placement.

Figure 4 compares the external flow cases with EHD jet in a quiescent environment. The momentum displacement is shown in absolute values as momentum is subtracted from the boundary layer in the counter-flow case and added to the boundary layer in the counter-flow. *Debien et al.* show similar momentum displacement results with a planar geometry at approximately $V_{AC}$ = 30 kV and $f_{AC}$ = 1.5 kHz at quiescent conditions [101]. The literature did not present the momentum comparison between the co- and counter-flow injections.

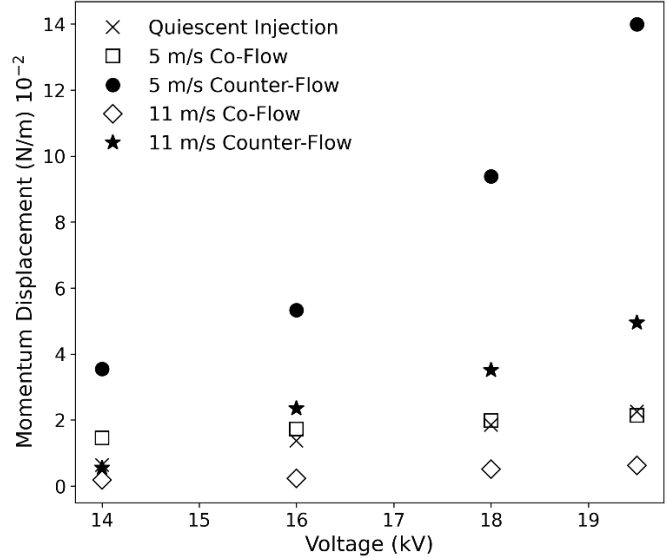

**Figure 13. DBD momentum displacement at $U_\infty = 5$ m/s and $U_\infty = 11$ m/s external flow**

In the co-flow scenarios, the momentum addition into the boundary layer is equal to or lower than the momentum of the EHD jet. The momentum change appears linear with $V_{AC}$ increase; however, the change in the momentum is relatively flat, suggesting that the momentum dissipation is driven primarily by the inner layer wall jet interaction with the wall. The 11 m/s co-flow cases show lower momentum addition at all voltages, likely due to viscous dissipation in a thinner boundary layer. Unlike the experiments in quiescent conditions [89], the fluid momentum of the EHD jet in the co-flow injection is not conserved as it travels downstream. In counter-flow configuration, the momentum displacement is more significant due to reverse flow at the surface. The highest displacement is observed in the counter-flow $U_\infty = 5$ m/s case. The momentum displacement is approximately 6.5x greater than its co-flow counterpart.

## 4. CONCLUSION

We have experimentally investigated the performance of a DBD plasma actuator over a range of voltages $V_{AC} = 12$ kV – 19.5 kV at 2 kHz in co-flow and counter-flow with freestream velocities of 5 m/s and 11 m/s. At lower free stream velocity, the EHD momentum injection increased velocity near the wall by ~ 2.0 m/s, effectively thinning the boundary layer. The results at $U_\infty = 11$ m/s show the diminishing influence of the DBD actuator. The counter-flow injection displaces up to 6.5 times more momentum than its co-flow counterpart at $U_\infty = 5$ m/s, resulting in the formation of a recirculation zone that grows with higher $V_{AC}$ voltages. At the higher speed setting, the DBD jet also displaces more momentum, but the effects are suppressed. Manipulation of this localized separation offers new opportunities for active flow control techniques.

The power usage associated with the DBD discharge is measured through capacitive measures with high temporal resolution through several cycles. There was no significant difference in power expenditure for all voltage and external flow conditions between the co-flow, counter-flow, and quiescent conditions. This experimental data set can be used for the development and validation of multiphysics models. Future research and development should extend to understanding and exploring the relationship between the unsteady forcing of the DBD and the turbulent characteristics of the external flow.

## 5. ACKNOWLEDGMENTS

This work was funded by the Joint Center for Aerospace Technology Innovation (JCATI).

# NOMENCLATURE

| | |
|---|---|
| $C$ | Pitot tube correction factor |
| $E$ | Electric field |
| $f_{AC}$ | Frequency of the applied voltage |
| $\vec{f}_{EHD}$ | Electro-hydrodynamic force term |
| $i(t)$ | Current |
| $I_{dis}$ | Discharge current |
| $L$ | Spanwise Length |
| $M$ | Momentum of the induced jet |
| $P$ | Pressure reading from the pitot tube |
| $W$ | Discharge energy consumption |
| $W_{mech}$ | Mechanical power |
| $W_{elec}$ | Electrical power |
| $U(y)$ | Velocity at y height |
| $U_{max}$ | Maximum velocity of the wall jet |
| $U_\infty$ | External flow velocity |
| $V_{AC}$ | AC Voltage in the DBD actuator |
| $v$ | Time-averaged velocity |
| $t^*$ | Normalized time value |
| $\rho$ | Density |
| $Q$ | Mass flow rate |